# POSITIVE AFFIRMATION OF NON-ALGORITHMIC INFORMATION PROCESSING

Carlos Eduardo Maldonado
Research Professor
School of Political Science and Government
Universidad del Rosario
ORCID: 0000-0002-9262-8879
carlos.maldonado@urosario.edu.co

**Abstract**

One of the most compelling problems in science consists in understanding how living systems process information. After all, the way they process information defines their capacities to learning and adaptation. There is an increasing consensus in that living systems are not machines in any sense. Biological hypercomputation (BH) is the concept coined that expresses that living beings process information non-algorithmically. Maldonado and Gómez (2015) have brought up (BH) as a new problem within complexity science. This paper aims at proving a positive understanding of "non-algorithmic" processes. A number of arguments are brought that support the claim. This fosters, it is argued, a brand new understanding of information processing among living beings. Some conclusions are drawn at the end.

Introduction

How do living systems process information? This problem has been stated as biological hypercomputation (BH) (Copeland, 2002; Stannett, 2006; Syropoulos, 2008; Stepney, 2009). A couple of very basic acknowledgements serve both as condition for BH and ground for it, thus:

i) Living systems are not machines, in any sense of the word
ii) Living systems do not process information as any kind of a TM (including Persistent Turing Machines, P-TM)

As a consequence:

iii) Living systems process information non-algorithmically

This conclusion deserves a thorough explanation, in order, little by little, to express this idea in a positive way. This means:

iii) Living systems process information non-algorithmically. But what does it properly, i.e. positively, non-algorithmically mean?

By contrast to the TM, and the strong Church-Turing Thesis (ChT-t) it means that living systems do not process information in terms of mathematical functions, rules, precepts, and the like. More exactly, they do not process information sequentially, linearly, or top-down.

Several approaches have been undertaken in order to better clarify that claim. Two conspicuous explanations concern emergent computation and interactive computation. We shall come back later to this. Being as it might be, it is my contention to solve the question in this paper.

Computationally speaking, nearly all current explanations and understandings about the information processing of the living can be synthetized in the following terms:

Petacapacity ---------------------------------------Clusters----------------------------------------e!

This equation reads: having a power of hundreds of petabytes, a cluster of processors gathered in hundreds and thousands as robust clusters, usually networked in parallel, work 24/7 connected to energy – working hence 24/7. A computational model needs be literally plugged in order to work. This is expressed by: e!, meaning energy constantly plugged and feeding the processor (CPU). From a biological standpoint this can easily be stated as if living beings were 24/7 conscious about their environment. Awareness, on one level, and consciousness, on another can be grasped as a state of being-plugged with the world and nature. Consciousness or awareness is here simply information processing (Dodig-Crnkovic, 2016). Anything else follows from that processing – after all, literally, life and death. We shall call such 24/7-petacapacity-clusters-e! situation as "*plugged"-state*.

Thanks to such characteristics –petacapacity, cliusters, e!-, gene sequencing is possible, the search for extraterrestrial bodies, among them black bodies, is carried out, and the spying of the Internet by the CIA, as reported recently, is worked out, among a number of other military and scientific endeavors.

The problem, though, is that no living being is "connected" ("plugged") 24/7. For instance, animals sleep, play, and are surprised by the environment, react differently to different stimuli or events. That could very much be said about plants (Baluska, 2009; Gagliano *et al*., 2010). Briefly said, living systems do mistakes, err, and happen to be – wrong. Many times, the outcome is that selection operates and fitness, i.e. adaptation is not possible, any longer. When an organism or a species do not process information rightly, the outcome is that it becomes endangered, and it might become endemic, and disappear eventually. Ontologically said, besides cognition and necessity, there is also chance and randomness. They cannot be discharged.

1-. <u>Biological hypercomputation, revisited</u>

The very basic assumption in and for (BH) consists in reckoning that living beings are not machines, in any sense of the word. Even though they are physical entities, they cannot be reduced just to physics, and certainly not to classical physics. Let us say that, to-date, we still do not know completely what matter is, and most probably quantum theory (not just quantum physics. Therefore, the interplay between quantum physics, quantum chemistry, quantum biology, and the technologies based on quantum principles and behaviors, mainly) may shed better lights in the future to come. Here, it can be sufficient to say that living beings cannot be explained mechanistically, in any concern. Computationally speaking, certainly not a TM – no matter what (U-TM, O-TM, etc.).

The most complex phenomenon in the known universe is life, i.e. living beings as-we-know-it, so much so that the sciences of life can be easily stated as life sciences event though the contrary cannot equally be said, namely that the life sciences are sciences of complexity. The distinction is made in science between life-as-we-know-it, and life-as-it-could-be (Rosen, 2000). The former refers to biology and the most basic characteristics of life is that it is humid, warm, and sticky, based on carbon with a DNA twisted to the right. The latter concerns artificial life, which is a research program aimed at understanding life, as such via the combination of software and hardware.

In sum, living beings process information non-classically (Dodig-Crnkovic, 2014). Even though living beings are partially classical they do not process information classically; i.e. like a machine – of any sort. Life is a phenomenon that cannot be compressed. Rather, it is understood and explained as it runs. The incompressibility of the information processing shows the process is open-ended, thus: relatively closed at the left, and open at the right, when seen as a "tape" that runs from left to right.

Living beings process information in a close intertwining with the environment – whether natural or cultural, for instance. The life of living beings happens as a permanent co-evolution between the environment and the organisms and species. The most suitable concept here is adaptive landscape, or also rugged adaptive landscape. Thus, the world or nature, are not passive or just conditional factors for the understanding of life, or for the very possibilities of the living beings. Nature and the living creatures are two different faces of one and the same token.

As a consequence, interactive computation can be safely said to be the gate that opens up the doors of/for (BH). This, however, is not to be taken necessarily as if interactive computation was already (BH). It properly means that the interactions among individuals, species, and with their environment constitute a condition sine qua non for the very way how living beings process information. In other words, living beings compute interacting, but the interaction cannot be taken in a reductionist way, for there are many other ways in which living systems process information.

Since life does not process information in terms of classical logical or mathematical functions, new mathematics and non-classical logics seem to be needed when understanding (BH). In mathematics, new developments are being undertaken, for instance in the direction of the harmonics, and automorphic functions (Frenkel, 2014). As for logics, some of the non-classical logics, in particular paraconsistent, epistemic and dynamic logics, seem to be very helpful when understanding life in terms different than formal classical functions.

Now, in computational terms the difference between software and hardware is irrelevant for the living beings. They are one and the same, hardware engineering and software engineering, form the cell level up to the ecosystems. No real distinction exists therein. From a biological point of view, epigenetics allow for an analogous claim.

Computing thereafter can safely be taken as metabolizing, i.e. transforming one thing into another. Now, the condition for metabolism is homeostasis. Consequently, computing rightly means for living beings making life possible, for a bad computation entails danger, error, and death. Processing information is to be grasped as (BH). Living beings process information (rightly), and live (well).

2-. A Problem

Traditionally, the explanation of information processing in human beings is viewed as taking place in the brain, and more exactly in the neocortex. Processing information is a conscious –aware and/or self-aware action that engages the various lobes of the superior functions – ultimately the symbolic functions that are determinant for a human-like understanding of the world and nature.

Consequently, artificial life, robotics and computational science have focused on the upper part of the brain. The outcomes are well known, such as Deep Blue, developed by IBM. Poker, chess, and go are the battlefields that strive for the search of a superior or higher intelligence.

In any case, the analogy or metaphor between computation and the brain is already old and well known. The analogy however forgets that the human brain is after all truly a set of three brains within one, namely, the reptilian brain (reptilian complex), the mammalian brain (limbic system), and the human brain (neocortex).

The trouble consists in that whereas much progress has been done concerning the simulation and developments of the human brain (= neocortex), little or none progress has been made concerning the simulation of the mammalian and reptilian brains. Computers, i. e., robots can barely identify their own temperature, emotions and feelings, for instance.

Deep Blue is not a minor achievement, no matter what. Humans have begun to be defeated in a number of challenges and problems. The story will continue, and artificial intelligence has a wide-open future in front of it. Yet, the difficulty of computationally understanding the two basic ancient brains challenges a sound understanding of information processing in artificial systems, a paradigm that seeks to bring insights about the human brain. Any computational approach that is not able to incorporate the two lower levels is indeed partial and limited.

The truth is that human beings, for example, do not think without the body, and do not thinking without emotions, and body self-awareness, either, for example. The reptilian complex and the limbic system remain as big challenges for the interplay between artificial intelligence and artificial life.

*En passant*: the ongoing program for terra-formation set to start for 2020 seems to forget that human beings are not possible without plants, as well as without animal, say, pets. Life is a large cooperative network that operates non-linearly, i.e., in unpredicted ways, many of them latent or tacit. Terra-formation without plants and animals is an endeavor destined to fail.

4-. Positive characteristics of non-algorithmic information processing: Random computation

Living systems do not *continuously* process information. In the case of human beings, psychology has already made it clear that most of time, human beings act as in automatic pilot (Horowitz, 2013). This means, they are not constantly and unceasingly reading and interpreting the world or the environment. Such is an over-intellectualized or over-rational statement – that corresponds to the Cartesian tradition.

Instead, living beings process information *discretely*. Living beings are discrete systems it follows – which has enormous consequences from many points of view. Here, I cannot go into details about the consequences of such an argument, and it must remain apart for another text to come. Some of reasons that explain that an animal, f.i., is not "conscious" all the time is that they play, err, make even mistakes, sleep, and the like.

Living beings process information as novelty and unpredictability happens, as the environment changes or another species or individual appears in her own surrounding, for instance. Straightforwardly said, living beings process information as it is needed, not before, not afterwards.

In other words, there is randomness. And randomness cannot be by any means predicted, and if "rationalized", it cannot be controlled. On the contrary, it fosters new types of computation and information processing previously unknown, or rarely activated.

Living systems – say, plants, animals, human beings – do not process information all the time. They are not Taylorian or Fordist machines. They do compute very well, but not always "perfectly". A Fordist or Taylorian machine is the one whose life-is-made-to-work, or also, it-is-life-made-thanks-to-work. And work demands of permanent conscious and aware attitudes. Shortly said, work demands that a living system be "plugged", as mentioned above.

Living systems create information about the world but at the same time change the information about the world. Short-term and long-term memories act and interplay with each other. By both creating and changing the information processed the very world is created and changed, indeed. In other words, there is no reality before the processing of information, but neither afterwards. More radically, there is no reality outside the information processing.

Table No. 1 provides a view of three different scenarios for properly understanding and solving (BH).

**Table No. 1: What Biological Hypercomputation Is Not, Could Be, and Is**

| **Is Not** | **Could Be** | **Is** |
|---|---|---|
| Turing Machine | Emergent Computation | Random Computation |
| Classical Computation | Interactive Computation | |
| Random Algorithms | Quantum Computation | |
| Event Computing | Cognitive Info-Computation | |
| | Morphological Computation | |

Source: Own Elaboration

Table No. 1 can be read as follows:

*Is Not* – expresses the possibilities that are simply discharged. A Turing Machine corresponds in computational terms to what in physics can be rightly grasped as classical systems and behaviors – which in terms of logic consists in the third excluded principle (= either a proposition is true or its negation is true).

*Could be* – the burden of the proof falls on the side of each probable computation that might positively affirm non-algorithmic information processing (Crutchfield, Mitchell, 1995). One serious difficulty about the could-be candidates for solving (BH) is that they take for granted the "plugged"-state. Such a condition cannot go without saying – simply because it is not true, as mentioned already above.

*Is*: This is the thesis of this paper, In order to grasp it better it is convenient to make something explicit, thus:

Living beings do not process information in just one way. They combine different types of computation: indeed, random algorithms, event computing, some types of TM such as U-TM, and interactive computation, among others. Any machine can process only one kind of computation in each event. That is not the case with living beings. As a consequence, living beings are capable of learning, and thereafter, of adaptation. Learning means having the possibility to combine or mix different types of computation, if needed. The more suitable such a combination is carried out, the better the adaptation.

To be sure, good metabolizing allows for discrete, and not continuous information processing. Health, a non-reflective experience – or also a pre-predicative experience - is the outcome of good information processing. Sometimes, though, good information processing is a matter of chance, luck, too, as it happens.

Random computation is conceptually closely related to quantum indeterminacy, and hence, to quantum randomness. Therefore, quantum computation can be taken to be the best gate toward random computation, under the proviso of the "plugged"-state. The right name then would be quantum random computation – something that operationally does not exist, as yet. This means that living beings can/are to be seen as quantum systems, a very radical conclusion. If true or tenable, the outcome is that the world is quantum, and the classical world is a limit.

Living beings process information randomly. We assist here to a strong distancing from the classical model of science based particularly on classical mechanics, on to biology taken as systems biology on to quantum science. Translated into the understanding of medicine, such a transition goes from the classical epidemiological model based on generalization and cultural overlapping on to transpersonal medicine; i.e. personalize medicine based on the lecture of the genome, and the steps that follow from it.

Conclusions

This is a propositive paper. It aims at saying that something else – and better – must be argued when claiming that living beings process information non-algorithmically. It goes without saying that non-algorithmic means non-linear, non-sequential, non-hierarchical, non-mechanical, for example, and that there is no canon about (information) processing, no matter what.

Can we think the "non" positively? This paper argues that it is possible. But in order to do that, we must reckon that living beings do not process continuously, but discretely. Moreover, living beings are not in a "plugged"-state. When seen from the standpoint of human beings, the information processing does not only and mainly take place in the neo-cortex, but it involves also the reptilian complex and the limbic system. To-date the main focus has been placed on the role of the neo-cortex. Good justifications for so doing can be readily been found in the literature. This paper argues that the information processing does involve the three brains at once, overlapped, if you wish. To-day we do not exactly know how they interplay for the processing of information. That is why biological hypercomputation consists in saying that living beings process information in or as random computation.